\newcommand{\review}{\iffalse} 

\review
\documentclass[preprint,review,12pt]{elsarticle}
\else
\documentclass[final,5p,twocolumn,times]{elsarticle}
\fi

\usepackage[utf8]{inputenc}
\usepackage{graphicx}
\usepackage{gensymb}
\usepackage{booktabs}
\usepackage{url}
\usepackage{lineno}

\iffalse
\usepackage{hyperref}\
\else
\usepackage{xcolor}
\definecolor{myblue}{rgb}{0.0, 0.0, 0.6}
\usepackage{hyperref}
\hypersetup {
  colorlinks = true,
  citecolor  = myblue,
  linkcolor  = myblue,
  urlcolor   = myblue
}
\fi
\usepackage{amssymb}
\usepackage{amsmath}
\usepackage{lineno}
\usepackage{setspace}

\journal{Nuclear Instruments and Methods in Research A}

\newcommand\h{${}^3$He\ }                   
\newcommand\hp{${}^3$He${}^+$\ }            
\newcommand\hpp{${}^3$He${}^{+\!\!\;+}$}  

\newcommand\hHe{${}^3$He${}^4$He\ }        
\newcommand\He{${}^4$He\ }                  
\newcommand\Ebeam{E_\text{beam}}
\newcommand\tCM{\theta_\text{CM}}
\newcommand\AN{A_\text{N}}

\begin{document}
\review \linenumbers \fi
\begin{frontmatter}

\author[1]{A.~Zelenski\;}
\author[1]{G.~Atoian\;\!}
\author[1]{E.~Beebe\;\!}
\author[1]{S.~Ikeda\;\!}
\author[1]{T.~Kanesue\;\!}
\author[1]{S.~Kondrashev\;\!}
\author[2,3]{J.~Maxwell\;\!}
\author[2]{R.~Milner\;\!}
\author[2]{M.~Musgrave\;\!}
\author[1]{M.~Okamura\;\!}
\review
\author[1]{A.\,A.~Poblaguev\;\!}
\else
\author[1]{A.\,A.~Poblaguev\;\!\corref{cor1}}\ead{poblaguev@bnl.gov}
\fi
\author[1]{D.~Raparia\;\!}
\author[1]{J.~Ritter\;\!}
\author[1]{A.~Sukhanov\;\!}
\author[1]{S.~Trabocchi\;\!}

\iftrue
\address[1]{Brookhaven National Laboratory, Upton, NY 11973, USA}
\address[2]{Laboratory for Nuclear Science, Massachusetts Institute of Technology, Cambridge, MA 02139, USA}
\address[3]{Thomas Jefferson National Accelerator Facility, Newport News, VA 23606, USA}
\else
\affiliation[1]{
            organization={Brookhaven National Laboratory}, 
            city={Upton},
            state={NY},
            postcode={11973}, 
            country={USA}}
\affiliation[2]{
            organization={Laboratory for Nuclear Science, Massachusetts Institute of Technology}, 
            city={Cambridge},
            state={MA},
            postcode={02139}, 
            country={USA}}
 \affiliation[3]{
            organization={Thomas Jefferson National Accelerator Facility}, 
            city={Newport News},
            state={VA},
            postcode={23606}, 
            country={USA}}
\fi
\review \else \cortext[cor1]{Corresponding author} \fi

\title{Optically Pumped Polarized \hpp\ Ion Source Development for RHIC/EIC}

\begin{abstract}
  \review \onehalfspacing \fi
  The proposed polarized \hpp\ acceleration in RHIC and the future Electron-Ion Collider will require about $2\times10^{11}$ ions in the source pulse. A new technique had been proposed for production of high intensity polarized \hpp\ ion beams.  It is based on ionization and accumulation of the \h gas (polarized by metastability-exchange optical pumping and in the 5\,T high magnetic field) in the existing Electron Beam Ion
  Source (EBIS).  A novel \h cryogenic purification and storage technique was developed to provide the required gas purity. An original gas refill and polarized \h gas injection to the EBIS long drift tubes, (which serves as the storage cell) were developed to ensure polarization preservation. An infrared laser system for optical pumping and polarization measurements in the high 3\:\!--\:\!5\,T field has been developed. The \h polarization 80\:\!--\:\!85\% (and sufficiently long $\sim$30\,min relaxation time) was obtained in the \lq\lq{open}\rq\rq\ cell configuration with refilling valve tube inlet and isolation valve closed. The development of the spin-rotator and \hHe absolute nuclear polarimeter at 6\,MeV \hpp\ beam energy is also presented.
\end{abstract}


\date{July 3, 2023}

\end{frontmatter}


\section{Introduction}

The Relativistic Heavy Ion Collider\,\cite{Ale2003} (RHIC) at Brookhaven National Laboratory (BNL) is the first high-energy accelerator-collider complex where the \textit{Siberian Snake} technique\,\cite{Derbenev:1978hv} has been successfully implemented to avoid the resonance depolarization during the beam acceleration and yields 60\% polarization for colliding beams. It also serves as the home for the future Electron-Ion Collider\,\cite{Accardi:2012qut} (EIC). 
The \hpp\ polarization can be preserved during acceleration in high-energy synchrotron accelerators like the Alternating Gradient Synchrotron and RHIC by using the \textit {Siberian snake} technique.   The nuclear polarization in a polarized \hpp\ beam is carried mostly by neutrons. Therefore, high-energy collisions of polarized electrons and neutrons can be effectively studied at EIC with the polarized \hpp\ beam.

The proposed polarized \hpp\ acceleration in RHIC will require about $2\!\times\!10^{11}$ ions in the source pulse and about 10$^{11}$ ions in the RHIC bunch.  To deliver this intensity in a 20\,$\mu$s pulse duration for the injection to the Booster, the source peak current has to be about 2000\,$\mu$A, which is $\times1000$ higher than ever achieved in previous \hpp\ ion sources.  We have proposed the concept for a polarized \hpp\ ion source based on the existing Electron Beam Ion Source (EBIS) at BNL\,\cite{Zel2003,Mil2011,Eps2013}. The \h atoms are polarized via the Metastability Exchange Optical Pumping technique\,\cite{Abb2004,Bat2005} in a glass cell at a pressure of 1\:\!--\:\!10\,mbar in a high-field 5.0\,T magnetic field within the EBIS solenoid and then will be injected into the EBIS ionizer. The ionization is produced in the 5.0\,T magnetic field. This field is much higher than the critical 0.31\,T field for \hp ions, therefore the depolarization  in the intermediate single-charged \hp states is strongly suppressed (to less than 1\%). Some polarization losses may occur during ionization in the low field near solenoid edges. These effects will be studied experimentally with the polarized \hpp\ beam. The simulations show that polarization losses should not exceed a few (3\:\!--\:\!5) percent and the nuclear polarization of the \hpp\ beam in excess of 70\% can be achieved in this source. 

In the EBIS, an estimated $(2.5\text{\:\!--\:\!}5.0)\times10^{11}$ \hpp\ ions can be produced and accumulated in the long 1.5\,m EBIS trap region. The total charge capacity is estimated at about 10$^{12}$ based on the total electron beam charge and a neutralization factor of about 0.5. The  required beam intensity of $2\!\times\!10^{11}$ \hpp/pulse can then be obtained during extraction and acceleration in a single beam pulse.

Successful tests of polarizing \h in a high magnetic field\,\cite{Maxwell:2016knr,Zelenski:2018umf,Zelenski:2018anj,Max2020} have led to the development of the Extended EBIS concept. This upgrade will also improve the heavy ion and gas species production.  The Extended EBIS upgrade is now an approved Accelerator Improvement Project at BNL with the primary purpose of increasing the Au$^{32+}$ intensity, but it will provide essential infrastructure for the polarized \h ion source.

\section{Polarized \hpp\ Ion Source}

\begin{figure*}[t!!]
\centerline{\includegraphics[width = 0.9\textwidth]{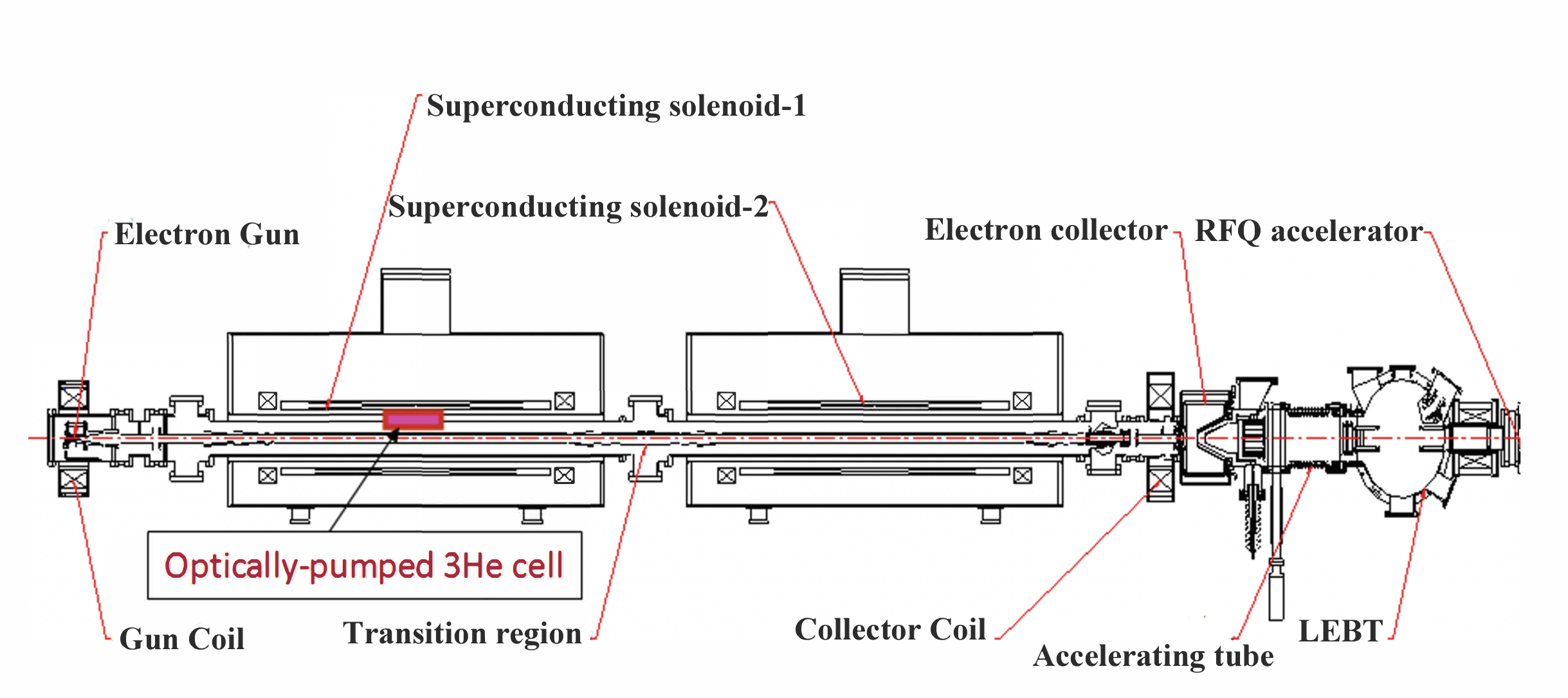}}
\caption{Schematic diagram of the extended EBIS. The polarized \h gas is injected into the drift tube of the new \lq\lq{injector}\rq\rq\ EBIS section. Here, RFQ means Radio Frequency Quadrupole and LEBT is the Low Energy Beam Transport.}
\label{fig:source}
\end{figure*}

The EBIS currently produces high charge state ions for injection to RHIC and will remain the primary source of charged ions from protons to Uranium for the future EIC. In the EBIS, the high intensity (10\,A) electron beam is produced by the electron gun with cathode diameter 9.2\,mm and injected into the 5.0\,T solenoid magnetic field. The electron beam is radially compressed by the magnetic field to a diameter of about 1.5\,mm in the ionization region and then expanded before dumping into the electron collector at the other end. Ions are radially confined by the space charge of the electron beam and longitudinally trapped by electrostatic barriers at the ends of the trap region. The ions are extracted by raising the potential of the trap and lowering the barrier\,\cite{Ale2010}. A second 5.0\,T solenoid has been constructed as an element in the extended EBIS upgrade. The polarized \h gas will be injected and ionized in the upstream solenoid, and \hp ions will be trapped and further ionized to the \hpp\ state in the downstream solenoid (see Fig.\,\ref{fig:source}). 

The \h gaseous cell will be placed inside the EBIS \lq\lq{injector}\rq\rq\ solenoid and the pulsed gas valve (similar to the valve in the BNL Optically Pumped Polarized Ion Source\,\cite{Zelenski:2018wrr}) will be used for the gas injection into the center of the EBIS drift tube system to minimize depolarization and increase ionization efficiency. The second \lq\lq{injector}\rq\rq\ EBIS section allows the use of differential pumping between the \lq\lq{gas injector}\rq\rq\ and the main EBIS. This is especially beneficial for gas species production (including \h gas). An isolation valve between the two EBIS sections will simplify the \h polarizing apparatus maintenance. The ionization in the EBIS is produced in a 5.0\,T magnetic field, which preserves the nuclear \h polarization while in the intermediate single-charged \hp state. The number of ions is limited to the maximum charge, which can be confined in the EBIS.  From experiments with Au$^{32+}$ ion production, one expects more than $2\!\times\!10^{11}$ \hpp\ ions/pulse to be produced and extracted for the subsequent acceleration and the injection into RHIC. After the \hpp\ beam acceleration to the energy 6\,MeV the absolute nuclear polarimeter based \hHe collisions will be used for the polarization measurements.

The high \h nuclear polarization in excess of 80\% was achieved by the metastability-exchange technique in the sealed glass cell in the high 2.0\:\!--\:\!4.0\,T magnetic field. In these measurements, the \h gas at 1.0\:\!--\:\!3.0\,Torr pressure was contained in the glass cell and the weak RF discharge was introduced to populate the metastable states.  Metastable atoms in the 2$^3$S$_1$ state were polarized by optical pumping with circularly polarized (2$^3S_1$\:\!--\:\!2$^3P_0$) 1083\,nm laser light. Any contamination in the helium gas cell (hydrogen, water vapor etc.) reduces the \h polarization due to metastable states quenching.  

\section{\h Gas Purification and Cell Filling System}

\review \begin{figure*}[t!!] \else \begin{figure}[t!!] \fi
\review
\centerline{\includegraphics[width = 0.45\textwidth]{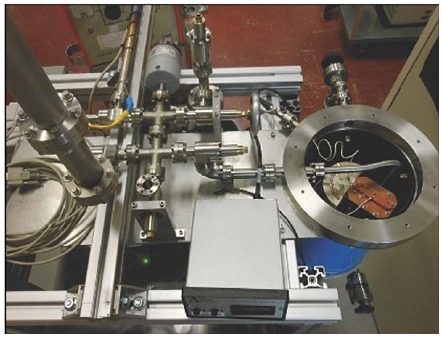}
            \hspace{0.07\textwidth}
            \includegraphics[width = 0.391\textwidth]{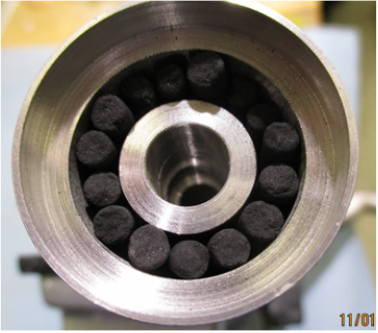}}
\caption{
  Left: the cryogenic \h purification and filling system. Right: the vessel filled with charcoal granules is attached to the cold head of the cryo-pump.
}
\else
\centerline{\includegraphics[width = 0.8\columnwidth]{Fig02a_cryo.pdf}}
\centerline{\includegraphics[width = 0.6\columnwidth]{Fig02b_cryo.pdf}}
\caption{
  Top: the cryogenic \h purification and filling system. Bottom: the vessel filled with charcoal granules is attached to the cold head of the cryo-pump.
}
\fi
\label{fig:cryo}
\review \end{figure*} \else \end{figure} \fi

The gas purity in the sealed cell was achieved by use of an elaborate glass cell cleaning, baking, outgassing procedure and sophisticated gas purification system. In the polarized source, the optically pumped cell must be connected to the valve for gas injection to the drift tube and the line for the gas refill. To eliminate contamination and to maintain the necessary gas purity in this \lq\lq{open cell}\rq\rq\ configuration we developed the system for \h gas purification and filling based on the technique of cryo-pumping, which pumps all gases except for the helium. In a conventional  two stage cryopump we cut away half of the cryo-panel (the second part is required to maintain the isolation vacuum) and installed the additional cold vessel (attached to the cold head of the cryo-pump) filled with charcoal granules (see Fig.\,\ref{fig:cryo}).

It was connected to the \h filling system by a thin vacuum bellows (see Fig.\,\ref{fig:cryo}). At the operational temperatures of 20\:\!--\:\!25\,K, the pump was continuously absorbing (and reducing the partial pressures of hydrogen, water, hydrocarbons, and argon) to the level below $10^{-7}$\,Torr. This pump absorbs also quite a significant amount of \h gas (about 100\,cm$^3$). The absorbed gas is released by the vessel heating with a built-in cartridge heater. This provides gas storage and supply for \h-cell operation at the optimal optical-pumping pressure of 3\:\!--\:\!5\,Torr. The pressure stability is maintained with the heater feedback system on the gas pressure measured by a Baratron MKS-626 pressure transducer.

\begin{figure*}[t!!] \else \begin{figure}[t!!] \fi
\review
\centerline{
  \includegraphics[width = 0.52\textwidth]{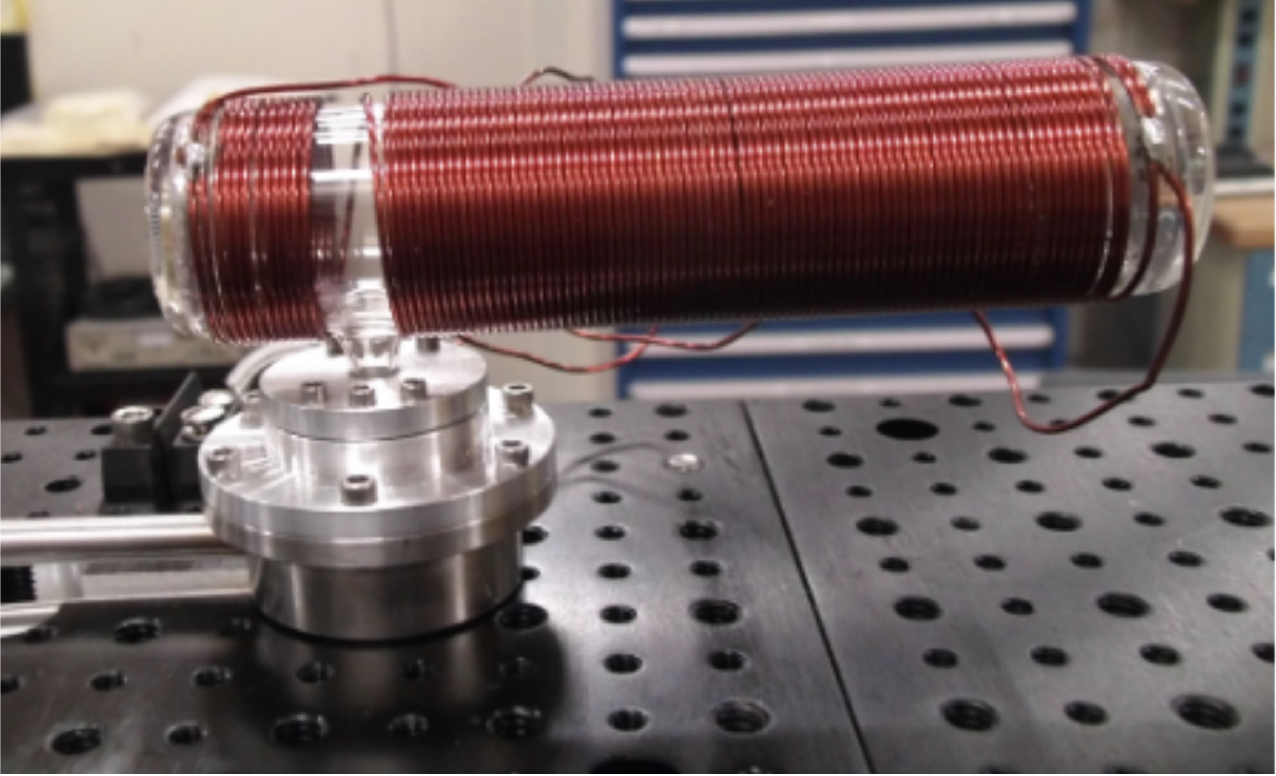}
  \includegraphics[width = 0.48\textwidth]{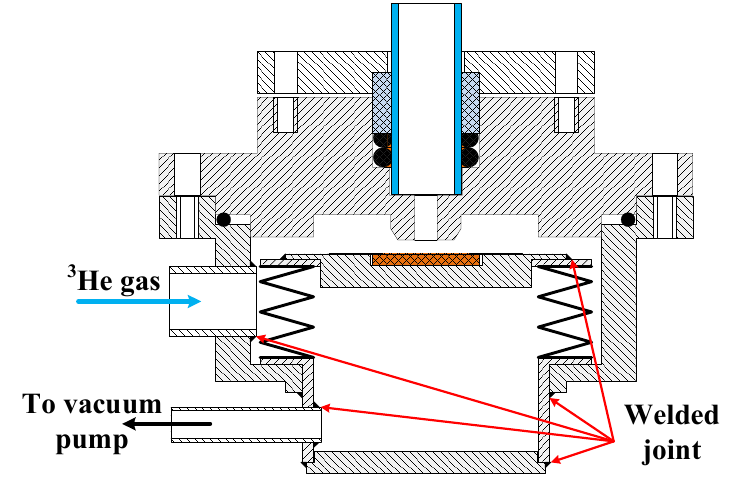}
}
\caption{
  Left:~\lq\lq{Open}\rq\rq\ \h gas cell (30 mm in diameter) with the isolation (filling) valve (IV) attached.
  Right:~A new custom built pneumatic (bellows-based) isolation valve.
}
\else
\centerline{\includegraphics[width = 0.9\columnwidth]{Fig03a_tvalve.pdf}}
\centerline{\includegraphics[width = 0.8\columnwidth]{Fig03b_tvalve.pdf}}
\caption{
  Top:~\lq\lq{Open}\rq\rq\ \h gas cell (30\,mm in diameter) with the isolation (filling) valve (IV) attached.
  Bottom:~A new custom built pneumatic (bellows-based) isolation valve.
}
\fi
\label{fig:cell}
\review \end{figure*}

The optically pumped \h glass cell is attached to the gas filling system with a 200\,cm long stainless tube. The cell and filling system were mounted on a movable support and inserted inside the superconducting solenoid. To prevent \h atoms depolarization due to travel through the solenoid gradient field we installed an additional isolation valve close to the cell in the homogeneous field region. Initially we used copper tube coupling and a commercial pneumatic-driven valve, which produced quite a bit of contamination and diffusion in the tube outside of the cell. As the latest improvement, we developed a new remotely controlled (pneumatic) valve with a small bellows.  The valve coupling to the He cell is designed to minimize the contact surface to aluminum surfaces and silicon sealing O-rings (see Fig.\,\ref{fig:cell}). To open the valve, it is connected to a vacuum pump, and it is closed by atmospheric pressure. This upgrade reduced \h gas contamination (as observed from discharge spectra) and should also reduce polarization losses due to diffusion outside of the cell.

\review  \else \end{figure} \fi

\begin{figure*}[t!!] \else \begin{figure}[t!!] \fi
\review
\centerline{\includegraphics[width = 0.5\textwidth]{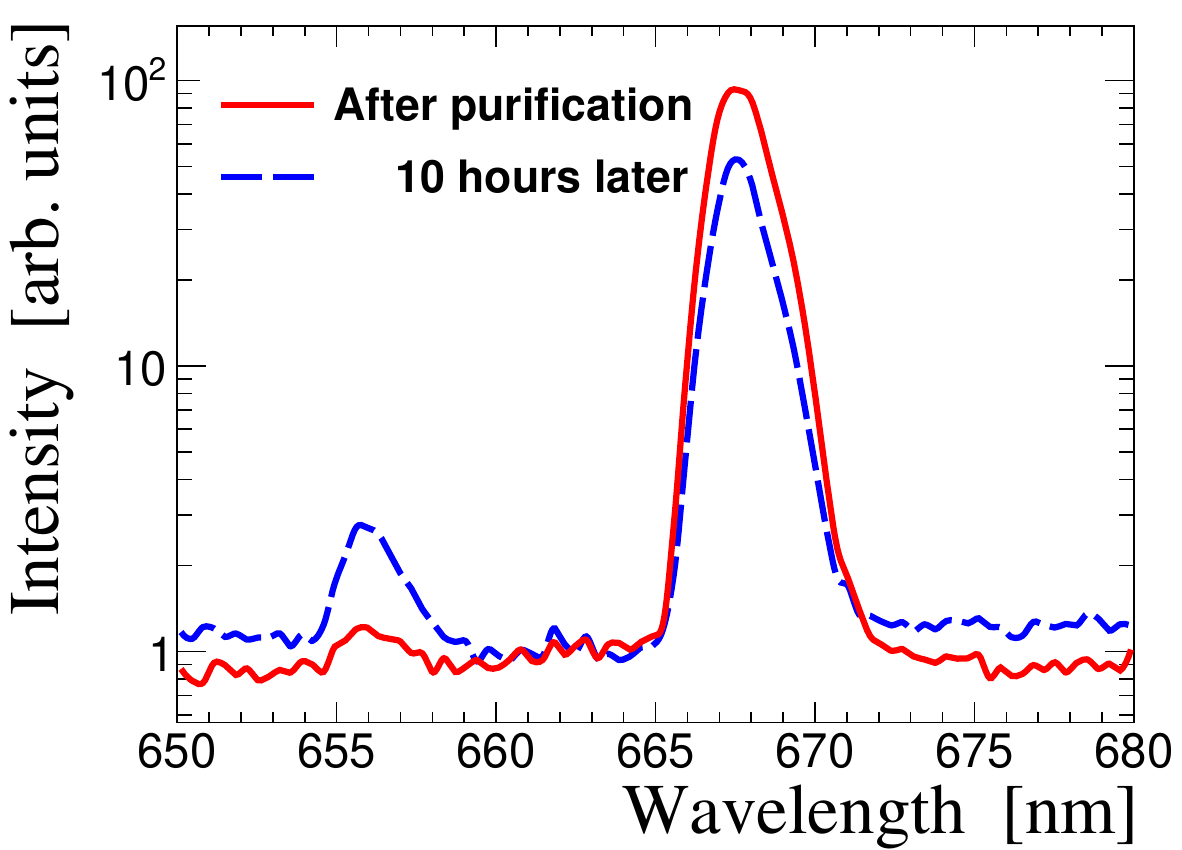}}
\else
\centerline{\includegraphics[width = 0.8\columnwidth]{Fig04_contamination.pdf}}
\fi
\caption{
\h contamination monitoring. The hydrogen Balmer-alpha line intensity at 656\,nm is strongly reduced, $<\!0.4\%$ (red solid line), relative to the 668\,nm \h line after the cryogenic purification cycle (cooling to 10\,K) and increased to $\sim\!3\%$ (dashed blue line)  after 10~h of RF discharge with the isolation valve closed.
 }
\label{fig:cont}
\review \end{figure*}

In the \h gas handling system, we employed all-metal bakeable valves, an oil free turbomolecular pump, and a residual gas analyzer (RGA). After extensive baking and pumping, the cryo-module is cooled down and the system is filled with about 100 standard cm$^3$ of \h gas, which initially is absorbed by the cryo-pump and then released by heating the vessel to about 20\,K to produce 3\:\!--\:\!5\,Torr He gas pressure in the cell and gas supply manifold. The RF-discharge in the \h-cell is induced by inductive coupling with the 100 turns coil (see Fig.\,\ref{fig:cell}). This inductive coupling works better than capacitive coupling for the cell in the high magnetic field. The inductive coupling produces a more homogeneous plasma density distribution across the cell volume (in contrast to the capacitive coupling where the discharge is mostly induced near the cell walls). We use the master oscillator and a 60\,dB broadband RF-amplifier to induce the discharge. The RF frequency is tuned for the best matching to the coil impedance. Typically, the RF-power is operated at about 44\,MHz frequency. The He gas purity in the RF-induced discharge in the cell is monitored by the measurement of the relative brightness of the hydrogen Balmer-alpha line at 656\,nm and adjacent \h spectral line at 668\,nm. With the cryo-pumping, the hydrogen contamination from residual gas equilibrium pressure in the cryo-pump, or water dissociation by RF-discharge in the cell is the main contamination and it is well monitored by the optical spectrometer. It was experimentally established that for production of high ($>\!80\%$) \h polarization, the relative hydrogen Balmer-alpha line brightness must be less than 2\% of the He-line. After the extensive He-filling system pumping, baking and \h cryogenic purification, the hydrogen line was almost completely eliminated (see Fig.\,\ref{fig:cont}) and high \h polarization can be maintained for several (3\:\!--\:\!5) hours with the isolation valve closed (which is required for high polarization production). This is sufficiently long for the polarized \h beam production in EBIS for the EIC fill.  

\review  \else \end{figure} \fi

\begin{figure*}[t!!]
  \centerline{%
    \includegraphics[width = 0.40\textwidth]{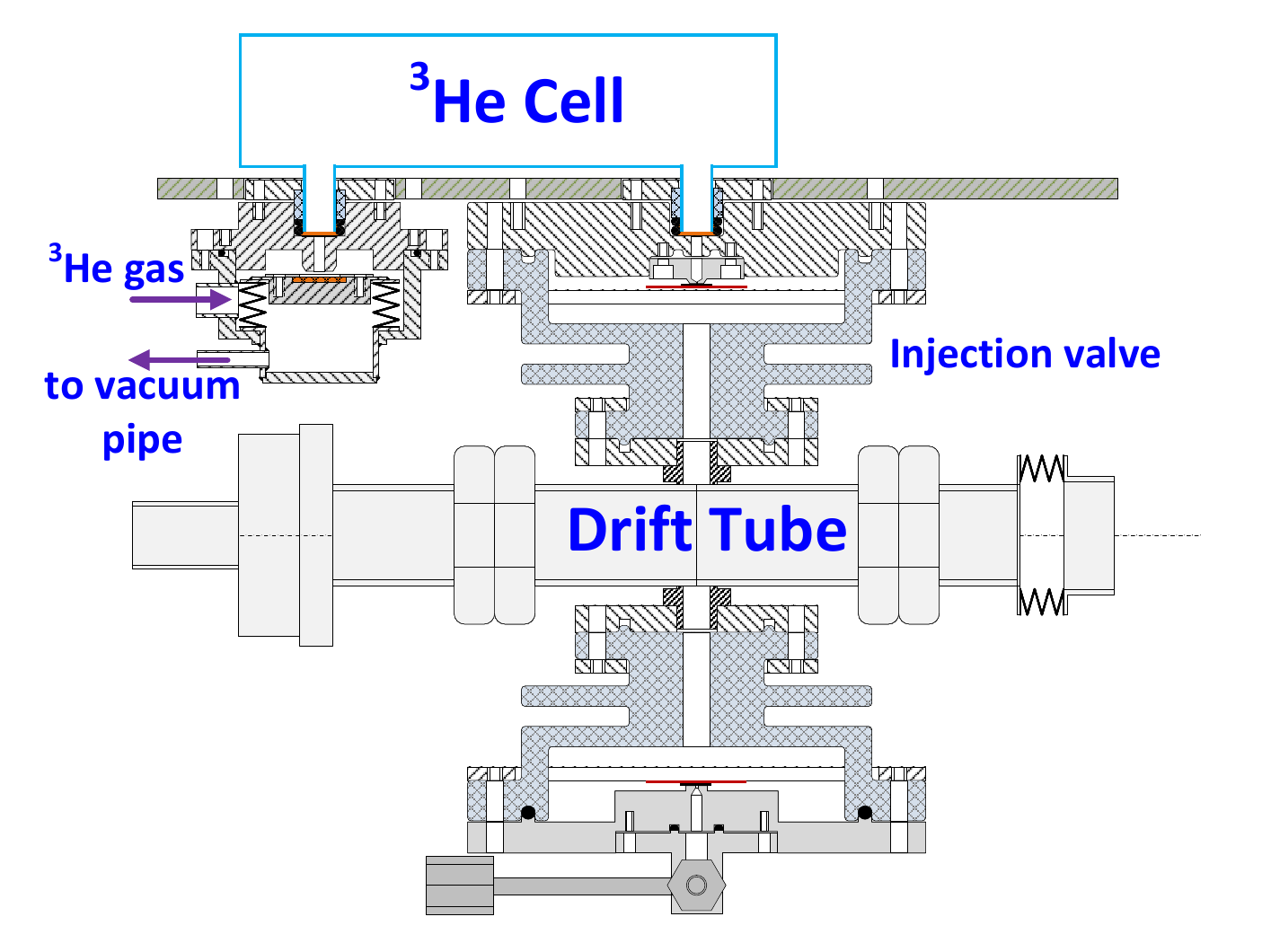}
    \includegraphics[width = 0.194\textwidth]{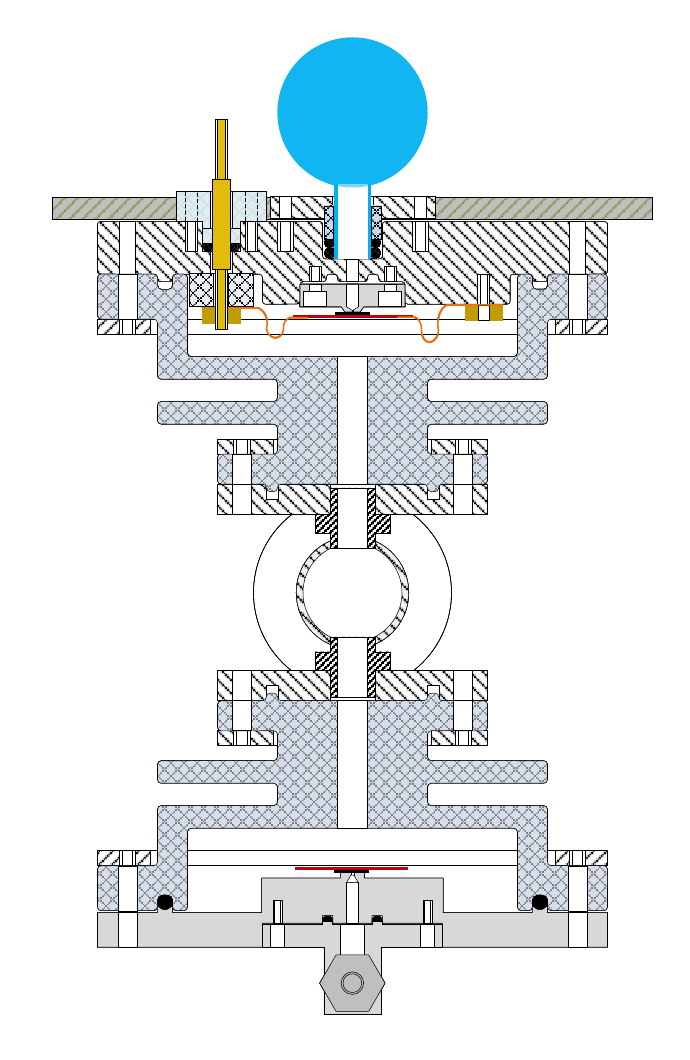}
  }
  \caption{Optically-pumped \h cell layout inside of the 5\,T EBIS solenoid.  The warm bore diameter is 215\,mm. Left: side view.  Right: end view.}
  \label{fig:pcell}
\end{figure*}

We have studied a new EBIS drift-tube configuration to increase the gas efficiency (minimize amount of injected \h gas for the EBIS trap saturation). The \h gas was injected into the small (inner diameter 10\:\!--\:\!20\,mm) drift tube by the pulsed valve. Estimates show that a very small amount of \h gas of about (5\:\!--\:\!10)$\times10^{12}$ atoms will be required to be injected into the drift tube for $\sim\!50\%$ EBIS trap neutralization. The total number of \h atoms in the 50\,cm$^3$ at 3\,Torr pressure is about $5\!\times\!10^{18}$ atoms and only a small fraction will be used for the EIC fill. After a few hours cell operation with the closed isolation valve, some contamination is produced due to outgassing by discharge. There is plenty of time between injections to the EIC (typically $\sim\!8$\,h) to recirculate the \h gas in the system by turning the heater off, to pump gas in the cryopump to capture all the contamination and re-filling the cell to the operational pressure. This entire cycle takes about 10\,min.

\section{The \h Cell Layout Inside the 5.0\,T EBIS Solenoid}

A prototype layout for the optically-pumped \h cell inside the 5.0\,T EBIS solenoid is presented in Fig.\,\ref{fig:pcell}.  The EBIS superconducting solenoid warm orifice diameter is 200\,mm and we have to place the optically-pumped \h cell, injection valve and HV separation insulator in this radially very limited space.  Fortunately, laser beams are transported in fibers and are quite compact. 

We developed a pulsed valve for the \h-gas injection into the EBIS drift tube, which operates in the 5.0\,T solenoid field. In this valve, the pulsed current of 10\:\!--\:\!20\,A passes through the flexible springing plate (made of phosphor bronze with a thickness of 0.12\,mm). The sealing Kalrez circular pad (5\,mm in diameter, 1.0\,mm thick) was attached to the plate. The induced Lorentz (Laplace) force: $\boldsymbol{F}\!=\!eL\,[\boldsymbol{I}\times\boldsymbol{B}]$ = 2\:\!--\:\!5\,N (for L=5\,cm long plate) bends the plate and opens the small (0.1\,mm in diameter) hole for the gas  injection into the drift-tube. The valve prototype was tested in the 2.5\,T solenoid field. A gas flow as low as $2\!\times\!10^{12}$ atoms/pulse was measured at 12\,A current through the plate. The valve was also operated with the four consecutive pulses 4\,ms apart, producing up to 10$^{13}$ atoms/cycle. This might be an optimal mode for the gas injection distributed over 20\,ms for the effective ionization by the EBIS electron beam, while limiting the injection gas cell pressure to $\sim\!10^{-6}\,\text{mbar}$.

\review \begin{figure*}[t!!] \else \begin{figure}[t!!] \fi
\review 
\centerline{\includegraphics[width = 0.45\textwidth]{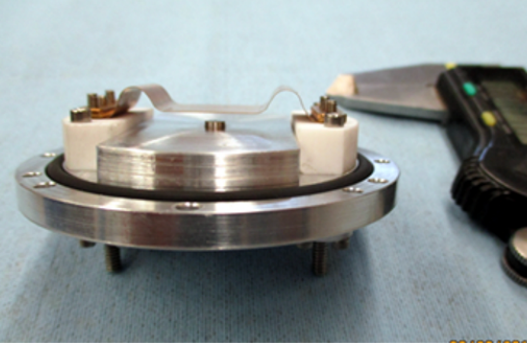}
            \hspace{0.10\textwidth}
            \includegraphics[width = 0.352\textwidth]{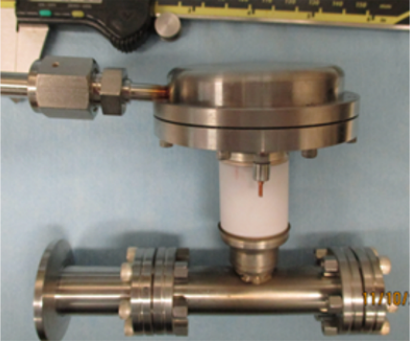}}
\caption{
  Left:~the $[\boldsymbol{I}\times\boldsymbol{B}]$ pulsed \h injection valve with the moving springing plate and two current feedthroughs.
  Right:~the injection valve for unpolarized gas injection, with the ceramic isolation section.}
\else
\centerline{\includegraphics[width = 0.7\columnwidth]{Fig06a_valve.pdf}}
\centerline{\includegraphics[width = 0.6\columnwidth]{Fig06b_valve.pdf}}
\caption{
  Top:~the $[\boldsymbol{I}\times\boldsymbol{B}]$ pulsed \h injection valve with the moving springing plate and two current feedthroughs.
  Bottom:~the injection valve for unpolarized gas injection, with the ceramic isolation section.}
\fi
\label{fig:valve}
\review \end{figure*} \else \end{figure} \fi 

The \h cell and gas preparation and filling system are at ground potential (actually at common HV EBIS potential of about 100\,kV). The drift tube potential is about 20\:\!--\:\!30\,kV. The \h cell and the pulsed injection valve are separated from the drift tube by a ceramic HV insulator. With the small amount of injected gas the pressure in the insulator is below $10^{-5}$\,mbar. In addition, the high 5\,T magnetic field reduces the ions transverse mobility which affects the discharge. We built a pulsed valve for the unpolarized gas injection into the EBIS using a two inches long ceramic insulator [see Fig.\,\ref{fig:valve}\,\review(Left)\else(Top)\fi]. It has been operated in the test EBIS setup by applying HV potential up to 40\,kV.

\section{\h Cell Optical Pumping and Polarization Measurements}

For tests of \h optical pumping at high fields, the 30\,mm \lq\lq{open-cell}\rq\rq\ was centered with the 152\,mm warm-bore of a 3.0\,T solenoid. To light the plasma discharge required to produce meta-stable states with the gas, a tightly-wound coil was driven with 44\,MHz RF, which was 50\% amplitude modulated at 297\,Hz. The pumping laser was a Keopsys continuous-wave, Ytterbium-fiber laser, which could provide 10\,W laser power at 1083\,nm wavelength with a 2\,GHz linewidth. This was tuned to pump the $f^+_4$ transition at 276.726 THz, and the laser frequencies were continuously monitored by WS-U wavemeter from HighFinesse. 

\begin{figure*}[t!!] \else \begin{figure}[t!!] \fi
\review \centerline{\includegraphics[width = 0.7\textwidth]{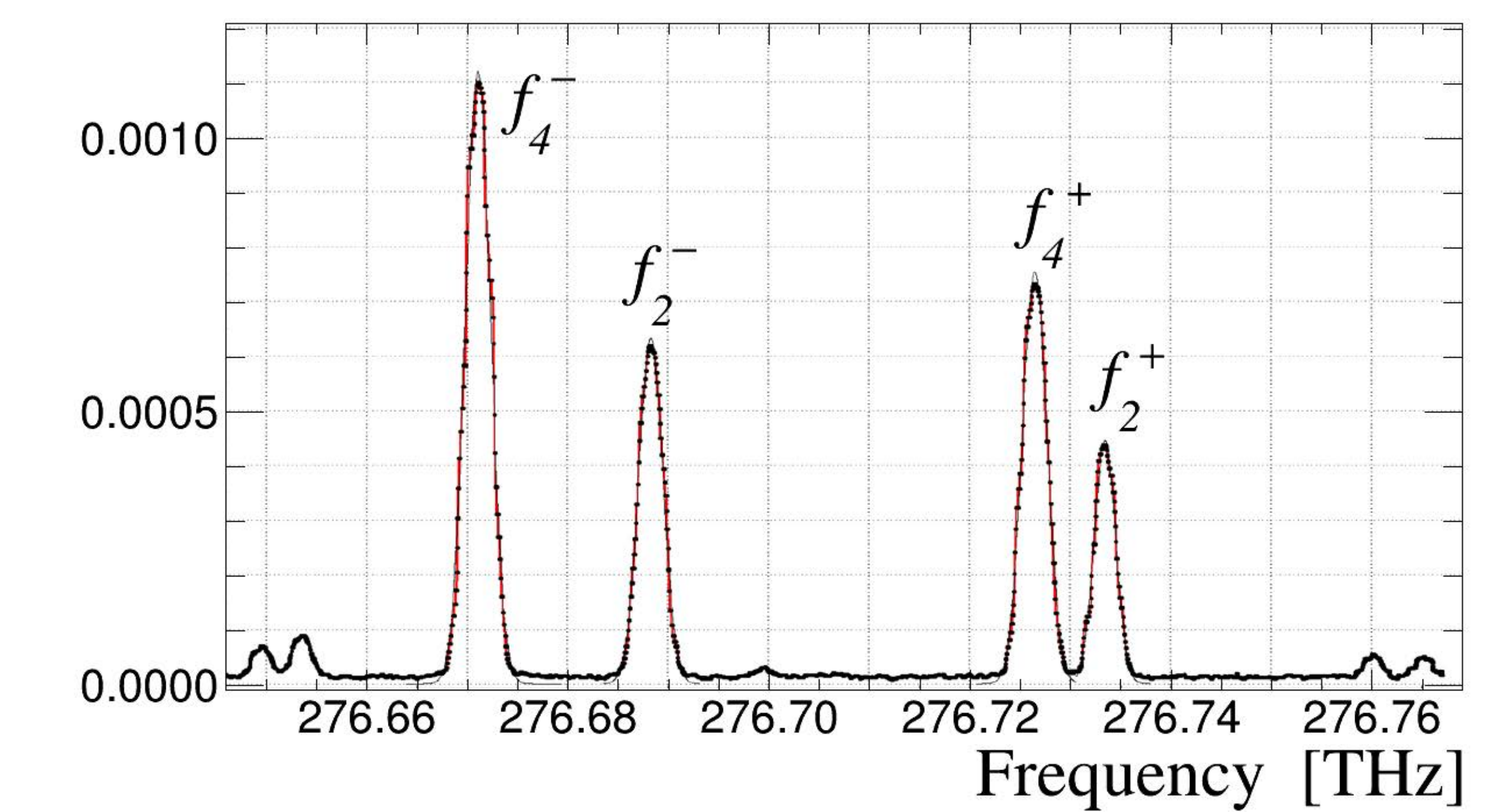}}
\else \centerline{\includegraphics[width = 0.95\columnwidth]{Fig07_abs.pdf}}
\fi
\caption{The \h absorption spectrum in a 2.0\,T magnetic field. The $f^+_{4-}$ transition at 276.726\,THz is used for optical pumping and a scan across transitions at 276.760\,THz is used for the polarization measurements.}
\label{fig:abs}
\review \end{figure*}

The polarization was measured via the absorption of a second probe laser \cite{Suc2007}, a technique which we have used in previous studies to measure near 90\% polarization with a 1\,Torr sealed cell between 2.0 and 4.0\,T \cite{Max2020}. A 70\,mW distributed feedback laser from Toptica was used to allow the rapid tuning of the laser wavelength via changes to the diode temperate and current. The probe laser passed through the pumping cell and was reflected through the cell a second time, before arriving at a photodiode, which allowed the monitoring of the absorption of the light as a function of laser wavelength.  Fig.\,\ref{fig:abs} shows a measured \h absorption spectrum at 3.0\,T, and includes the probe peaks used for these measurements at 276.76\,THz. The ratio of the peak-heights of the transition pairs, $r\!=\!a_2/a_1$, allows the calculation of an absolute measure of the nuclear polarization when calibrated at zero polarization: $M\!=\!(r/r_0-1)(r/r_0+1)$\,\cite{Suc2007}.

\review  \else \end{figure} \fi

\begin{figure*}[t!!] \else \begin{figure}[t!!] \fi
\review
\centerline{
  \includegraphics[width = 0.45\textwidth]{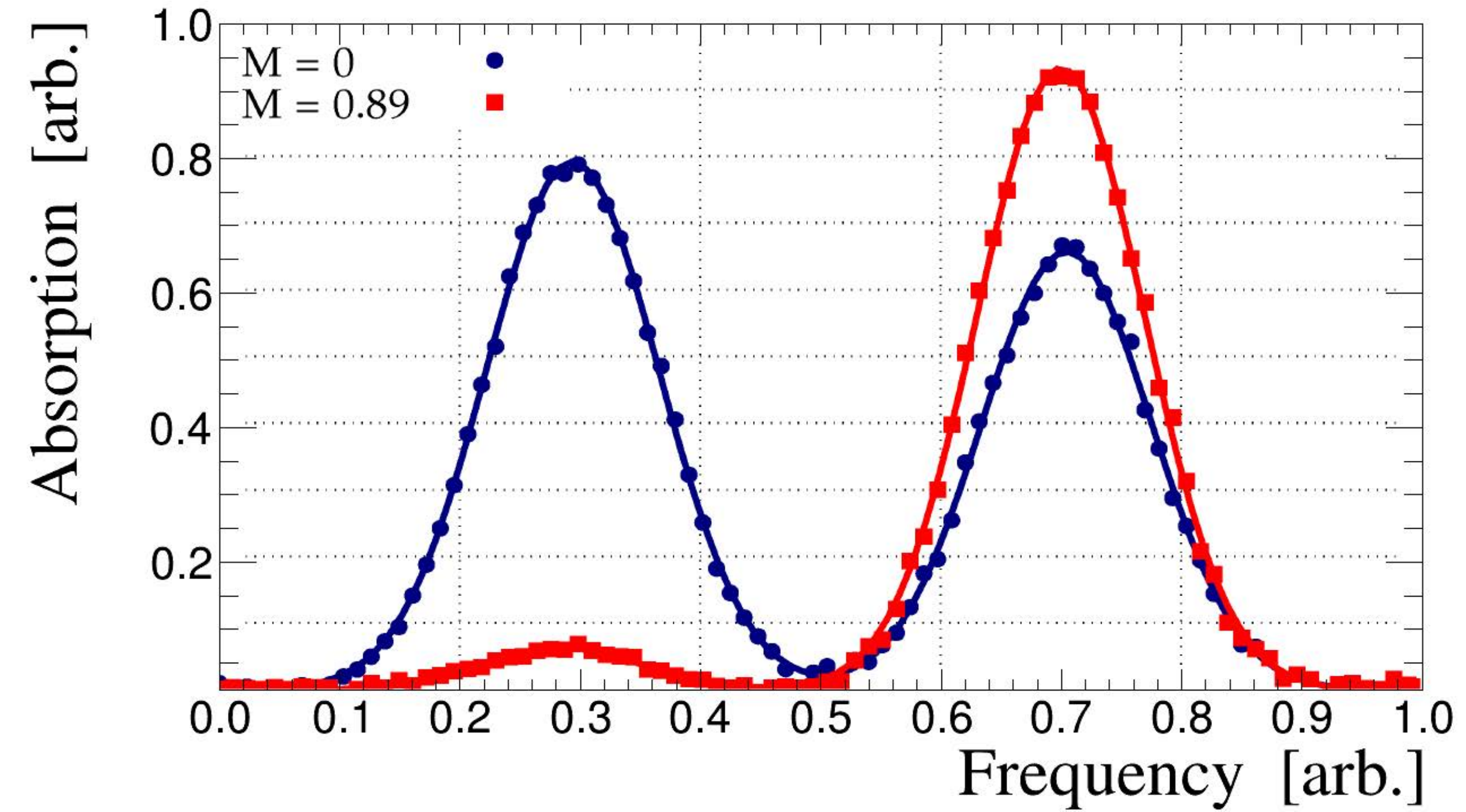}
  \hspace{0.05\textwidth}
  \includegraphics[width = 0.45\textwidth]{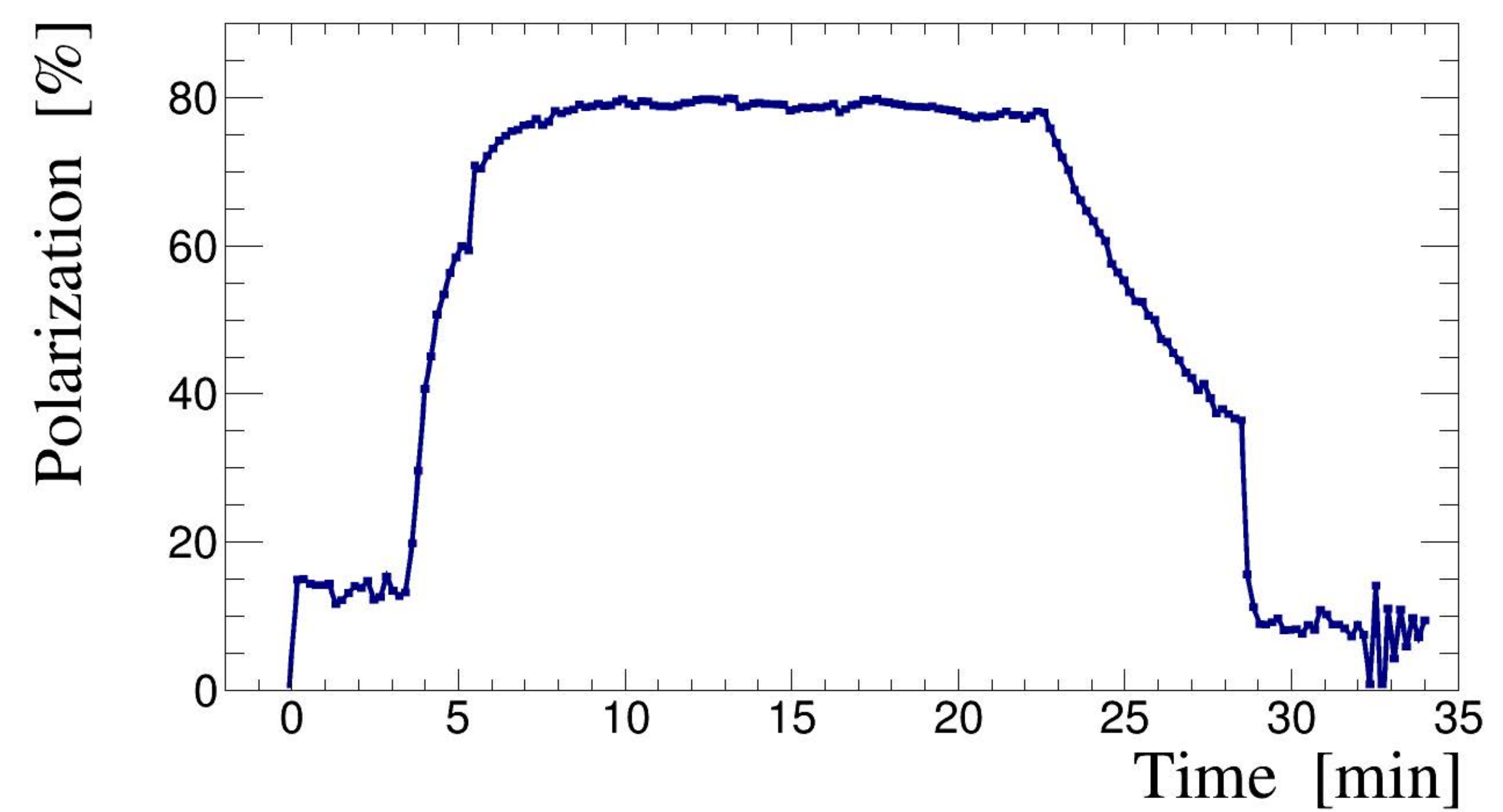}
}
\caption{
  Left:~example of the probe laser absorption signal for the 89\% polarization in a sealed \h cell at 3.0\,Torr pressure, compared with a signal at 0\%.
  Right:~polarization build-up vs. time. First 4\,min the isolation valve is opened (13\% polarization), after the valve was closed polarization increased to 80\% in 5~min, at 22\,min. the laser was switched off, and at 29\,min. the isolation valve was opened.}
\else
\centerline{\includegraphics[width = 0.9\columnwidth]{Fig08a_example_data.pdf}}
\centerline{\includegraphics[width = 0.9\columnwidth]{Fig08b_polarization.pdf}}
\caption{
  Top:~example of the probe laser absorption signal for the 89\% polarization in a sealed \h cell at 3.0\,Torr pressure, compared with a signal at 0\%.
  Bottom:~polarization build-up vs. time. First 4\,min the isolation valve is opened (13\% polarization), after the valve was closed polarization increased to 80\% in 5~min, at 22\,min. the laser was switched off, and at 29\,min. the isolation valve was opened.}
\fi
\label{fig:probe}
\review \end{figure*}

The amplitude modulation of the RF discharge at 297\,Hz was used to improve the sensitivity of the measurement by passing the photodiode signal to a lock-in amplifier. This allowed the isolation of only the portion of the signal which came from interaction with the discharge, reducing noise from the photodiode and other light sources. The polarimeter DAQ was integrated into the common RHIC control system, and the software allowed the display of the spectrum as well as Gaussian fits to each probe peak to determine peak height while reducing the effects of noise. An example of  measurement is presented in Fig.\,\ref{fig:probe}.
 
\review  \else \end{figure} \fi

The best results on optical pumping of \h gas in the \lq\lq{open}\rq\rq\ cell were $25\!\pm\!5\%$ with the open isolation valve and $80\!\pm\!5\%$ with the closed isolation valve at 3.0\,Torr pressure. In these measurements we used a preliminary gas cryogenic purification system, with operational temperature $\sim\!36\text{\:\!--\:\!}40$\,K (due to limited cooling power and higher heat transfer through the connecting tube). We also used a conventional brass Swagelock pneumatic isolation valve with the copper tube adapter to the \h cell. The gas purity has since been improved with the new cryogenic system and new isolation valve described above. We plan to continue optical pumping studies with the new improved gas preparation system.

\section{Chicane for Spin Rotation}

\begin{figure*}[t!!]
\centerline{\includegraphics[width = 1\textwidth]{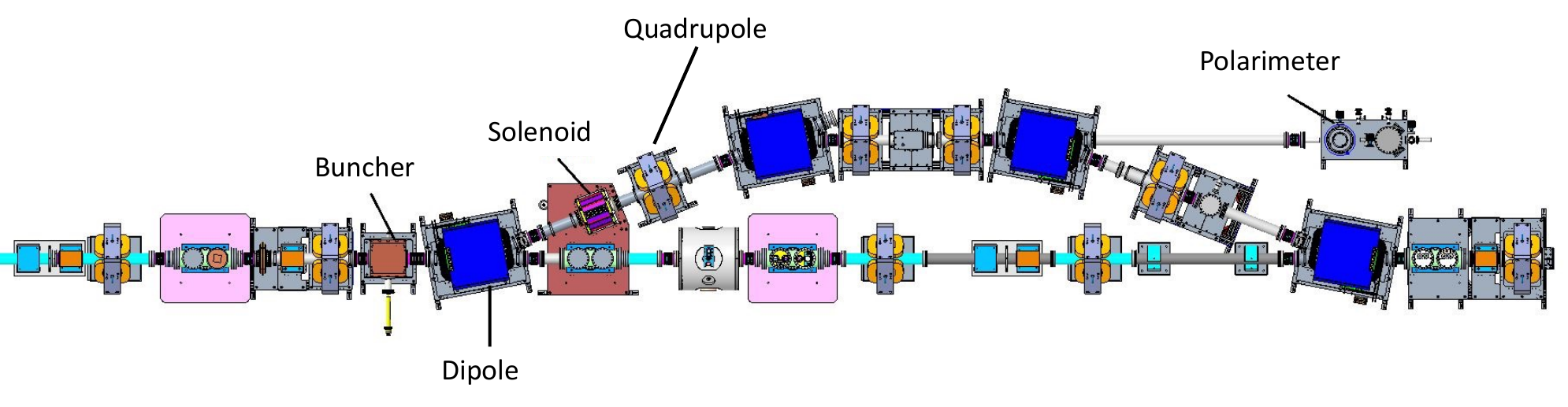}}
\caption{3D drawing of the chicane for the spin rotation. The trapezoidal chicane will be attached to the existing straight line between the EBIS injector and the Booster synchrotron. The spin rotation will be accomplished by the first dipole and spin flip by the following solenoid.}
\label{fig:chicane}
\end{figure*}

Polarized \h ions extracted from EBIS will be accelerated by two linear accelerators to 2\,MeV per nucleon.
The spin direction will be parallel to the beam line and will need to be rotated to the vertical direction before the injection into the Booster synchrotron.
The change of the spin direction will be done by the combination of a dipole and a solenoid. In the dipole field, the ion trajectory is deflected by $\theta$ = 21.5\degree. The dipole file also causes the spin to rotate with respect to the trajectory by 90\degree in the horizontal plane. Then the ion with the horizontally directed spin (perpendicular to the ion momentum) passes through the solenoid in which the spin is rotated around the solenoid axis by angle $\pm$90\degree depending on the polarity of the solenoid field. Thus, the spin direction is changed to vertical, upward or downward. The deflected ion trajectory is returned to the straight beam line by the following three dipoles. The beam may optionally be steered straight to a polarimeter rather than returned to the Booster injection line. The overall beam line will form a trapezoidal chicane as shown in Fig.\,\ref{fig:chicane}.

The beam optics design was done with TRACE\,3-D\,\cite{Crandall:1990fs} and confirmed by TraceWin\,\cite{Uriot:2015keg}.
The chicane was designed to be approximately symmetric to make it achromatic to minimize an emittance growth.
The optics simulation showed that a 5\,mA \hpp\ beam can be transported to the booster synchrotron and fit the transverse acceptance of the injection electrode with a required momentum spread ($\pm0.5\%$).

A RF buncher is placed in front of the chicane to reduce the energy spread of the ions.
The position was determined to have an appropriately long bunch width and a symmetric chicane.
The effective voltage is 40\,kV.
The structure is a quarter-wave resonator with two gaps.
The outer size is 335\,mm $\times$ 351\,mm $\times$ 580\,mm.
The bore is 80\,mm in diameter and the gap length is 5\,mm.
The required RF power is 500\,W.

The four identical dipoles are positioned at the vertices of the chicane.
Two of them on the existing straight line are driven by a pulsed power supply to be turned on for \h ions and off for other ion species.
The dipole shape is rectangular and the outer size is 536\,mm\,(H) $\times$ 364\,mm\,(V) $\times$ 538\,mm\,(L).
The pole width and the gap are 260\,mm and 110\,mm, respectively.
The pole shim was designed to have a good field region within a diameter of 60\,mm.
The dipole is laminated for pulsed operation.
The field is 0.19\,T and the curvature of the trajectory is 1.6\,m.
48-turns coils are wound around the upper and lower poles.
The total current in each coil is 8.4\,kA turn.
The conductor has a 6.6\,mm by 6.6\,mm square cross-section with 4\,mm diameter cooling channel.
The inductance and the resistance will be 30\,mH and 0.12\,ohm.
A trim coil is wounded for the 2\% field adjustment.
The dipole was fabricated and the field was checked.
The field strength was as designed.
A significant difference of the field in ramp-up and ramp-down was not observed.

The solenoid will generate the integrated magnetic field strength of about 0.15\,T$\cdot$m for 90 degree spin rotation.
The field at the middle will be 0.66\,T.
The design is the same as used in the existing line.
The inner diameter and the length of the coil are 145\,mm and 202\,mm, respectively.
1\,kA will flow in a 9.5\,mm $\times$ 9.5\,mm conductor with 6.3\,mm cooling channel wound by 120 turns. The charge will be provided from a 9\,mF capacitor bank. The charging time of the capacitor bank is about 0.2\,s. The current waveform will be half-sine with the width of 12\,ms. A bridge circuit with 4 Silicon Controlled Rectifiers (SCR) at the output will be used to reverse the current direction. 2 SCR on one pair of the diagonal lines will be on for one current direction while the other 2 SCR on the other pair will be on for the other current direction. The polarity of the solenoid can be changed for every pulse. The solenoid and the pulsed power supply were built. The field was generated as designed and it was confirmed that the power supply can work with 5\,Hz at 1\,kA peak.

To transport the beam, four quadrupoles and three steering magnets are installed.
A Faraday cup at the end of the second dipole and a beam profile wire monitor between the second and the third dipoles will be used.

\section{Absolute Nuclear Polarization Measurement at 6 MeV}

\begin{figure}[t!!]
\review \centerline{\includegraphics[width = 0.7\columnwidth]{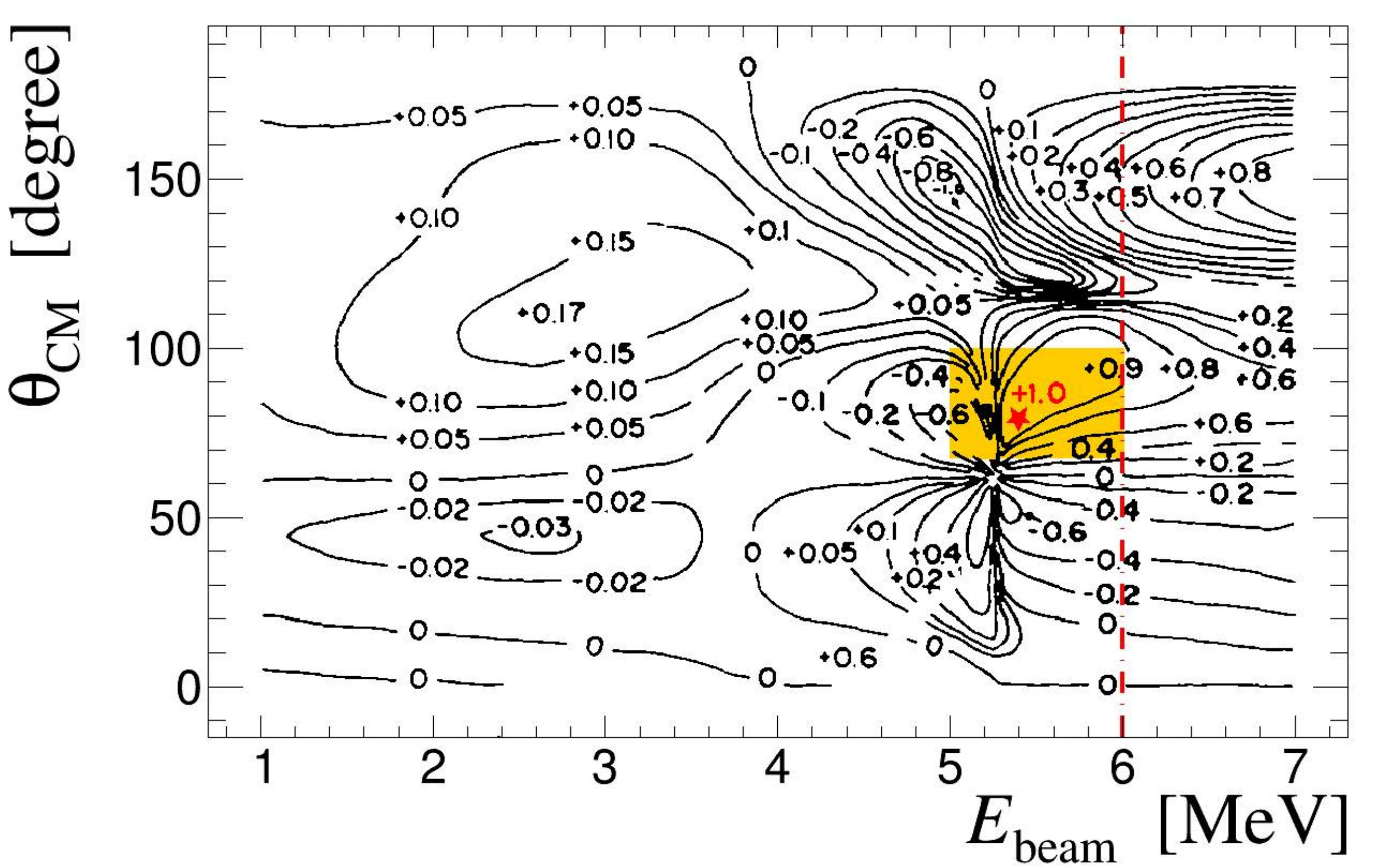}}
\else   \centerline{\includegraphics[width = \columnwidth]{Fig10_apower.pdf}}
\fi  
\caption{Polarization analyzing power in \hHe elastic scattering versus the \h beam energy $\Ebeam$ and the center-of-mass scattering angle $\tCM$ \cite{Boykin:1972can}. The expected location, $\Ebeam\!\approx\!5.4\,\text{MeV}$ and $\tCM\!\approx\!79\degree$, of the absolute maximum $A_N\!=\!+1$ is shown by red star marker. The area proposed for experimental calibration of measured analyzing power at $\Ebeam$=6\,MeV  (dashed red line) is shown by orange color.
}
\label{fig:apower}
\end{figure}

To directly verify polarization of the fully stripped \hpp\ ions extracted from the EBIS, a spin asymmetry measurement in scattering from unpolarized \He gas will be carried out. In Ref.\,\cite{Plattner:1971yej}, it was shown that for spin-1/2 particles scattered from particles without spin, analyzing power $A_\text{N}(E_\text{beam},\theta_\text{CM})$, as a function of the beam energy $E_\text{beam}$ and the center-of-mass scattering angle $\theta_\text{CM}$, must reach absolute maximum $|A_\text{N}|=1$ at some point $(E_\text{beam},\theta_\text{CM})$. To find the maximum, one can analyze experimentally determined phase shifts in the scattering. Using the experimental data\,\cite{Hardy:1970jkp}, several such points were established\,\cite{Plattner:1971yej,Boykin:1972can} for \hHe elastic scattering including one (Fig.\,\ref{fig:apower}) in the energy range of the EBIS Linac. However, actual accuracy of the available experimental data does not allow to precisely predict the point location.

For 6\,MeV polarized \h beam, the analyzing power has a local maximum, $\AN(6\,\text{MeV},96\degree)\!>\!0.9$, at $\tCM\!\approx\!96\degree$. By measuring the spin correlated asymmetry $a_\text{N}(T_h)=P_\text{beam}\AN(\Ebeam,\tCM)$ as a function of the scattered \h kinetic energy $T_h$ (kinematically, $T_h$ is strictly correlated with $\tCM$), one can determine the local maximum $a_\text{N}^\text{max}(\Ebeam)$ (at some value of $T_h^\text{max}$ or, equivalently, $\tCM^\text{max}$). Scanning the beam energy in the $5\!<\!\Ebeam\!\le\!6\,\text{MeV}$ range, the absolute maximum $a_\text{N}^\text{max}(\Ebeam^\text{max})=P_\text{beam}$ (at some beam energy $\Ebeam^\text{max}\approx5.4\,\text{MeV}$) can be found. Consequently, we can calibrate analyzing power,
\begin{equation}
    \AN(\Ebeam\!=\!6\,\text{MeV},T_h^\text{max}) = \frac
    {a_\text{N}^\text{max}(6\,\text{MeV})}
    {a_\text{N}^\text{max}(\Ebeam^\text{max})},
\end{equation}     
for the 6\,MeV beam energy. To implement the calibration method, the following should be considered:

--\quad%
The \h beam polarization must be stable during the calibration. To verify the stability, multiple measurements with each $\Ebeam$ can be used;

--\quad%
The scattered \h detector acceptance should be sufficiently large to isolate a local maximum of the asymmetry for any beam energy in the $5\!<\!\Ebeam\!\le\!6\,\text{MeV}$ range;

--\quad%
Two methods of variation of the \h beam energy are currently under consideration: {(i)} adjustment of the beam acceleration and {(ii)} changing the thickness of the polarimeter entrance window and, consequently, diluting the beam energy due to the $dE/dx$ losses.

Also, it is important to note that for the \h beam energy $\Ebeam\!\le\!6\,\text{MeV}$, there is no inelastic contributions to the \hHe scattering. 

\section{The \hHe Polarimeter Layout}

\begin{figure*}[t!!]
\centerline{
  \includegraphics[width = 0.8\textwidth]{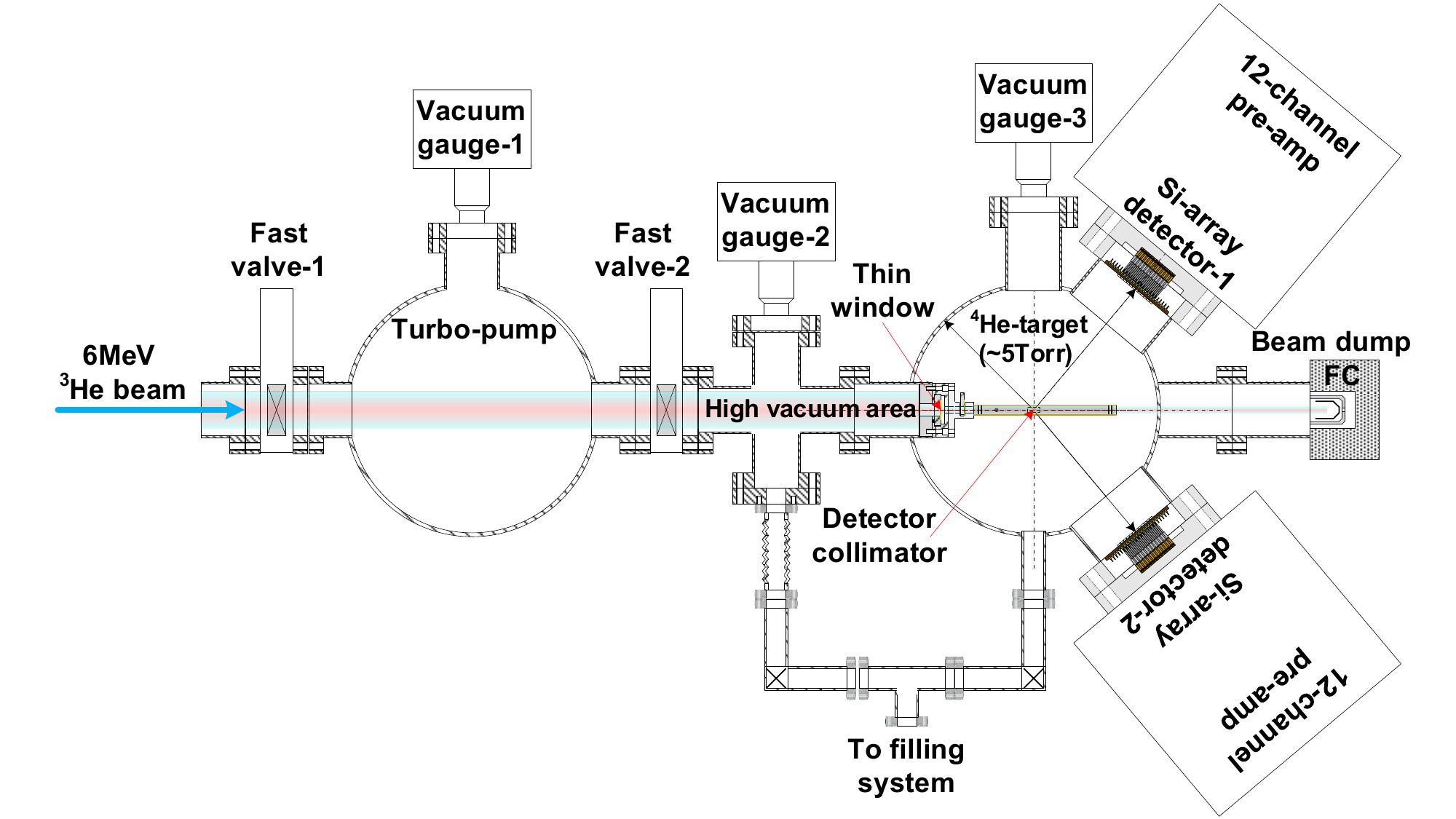}}
\caption{A layout of the \hHe polarimeter.  The effective target is a 1\,cm long collimated part of the 5\,Torr \He gas filled scattering chamber. The scattered \h and recoil \He angles, energies and time-of-flight are measured by the Si-strip detectors.}
\label{fig:pol}
\end{figure*}

The polarimeter layout is shown in Fig.\,\ref{fig:pol}. The polarized \h beam enters the scattering chamber through the thin aluminum foil window to minimize beam energy losses. 

The scattering chamber is filled with 5\,Torr \He gas. The effective target length of about  $\sim\!1\,\text{cm}$ is constrained by the collimators. Two Si detectors are located in the chamber at $\theta_\text{Lab}\!=\!\pm50\degree$, 10\,cm from the \lq\lq{center of the target}\rq\rq.  The expected displacement $\sim0.2$\,mm of the scattered/recoil particles in the detector due to multiple scattering is small compared to the Si strip width.  

The requirements on the detector geometry and measured energy range can be satisfied by Silicon strip detectors. For preliminary evaluation of the polarimeter performance we used the available Hamamatsu Si-photodiode array S4114-35Q\,\cite{S414-35Q}. Such a detector will cover center-of-mass angles $69\degree\!<\!\theta_\text{CM}\!<\!100\degree$. The detected particles energy range is 2.6\:\!--\:\!4.2\,MeV for \h and 1.5\:\!--\:\!3.4\,MeV for \He.  The 30 $\mu$m depletion region is sufficient for stopping 5.5\,MeV \h and 5.8\,MeV \He. The proposed design allows us to strongly suppress systematic errors. To this end, we can use many constraints resulting from the measured energy $T$, angle (the strip number) $\theta$, and time-of-flight for both, scattered \h and recoil \He. Detailed description of the \hHe scattering kinematics in the measurements is considered in Ref.\,\cite{Atoian:2020hny}. 

For energy calibration we will use $^{148}$Gd (3.183 MeV) and $^{241}$Am (5.486 MeV) $\alpha$-sources. Both dead-layer and gain can be determined. This method was proven well in the RHIC polarimeters\,\cite{Poblaguev:2020qbw}.

The expected energy resolution is $\sigma_E/E\!\le\!2\%$ and time resolution is $\sigma_t\!\le\!0.2\,\text{ns}$. The Si strip structure (1\,mm step) potentially provides a measurement of the scattering/recoil angle (lab. system) of about $\sigma_{\theta}\!\sim\!0.2\degree$. However, taking into account the effective size of the target ($\sim$1 cm), the effective angular resolution is about $\sigma_{\theta}\!\sim\!1.2\degree$, which corresponds to an effective $\sigma_E^\text{geom}\sim0.1\,\text{MeV}$. 

The detector is equipped with a standard 12-channel preamplifier and shaper from the RHIC p-Carbon\,\cite{Huang:2006cs} and Hydrogen Jet Target polarimeter\,\cite{Zelenski:2005mz}  (rise time $\sim20$\,ns and signal width $\sim60$\,ns).  For Data Acquisition we plan to use the VME 250\,MHz 14-bit waveform digitizers (SIS3316-250-14). Recording the full 20\,$\mu$s\,/\,5000\,samples, the bunch signal in every readout channel is essential for monitoring the possible rate dependent systematic errors. The expected data rate in the polarization measurements is $\sim10$\,kB/sec per readout channel. 

The preliminary evaluation of the polarimeter performance was based on the following assumptions about the \h beam: 6\,MeV energy, 70\% polarization, $5\!\times\!10^{11}\,\text{s}^{-1}$ intensity, 1\,Hz beam pulse repetition rate, and $\sim20\,\mu\text{s}$ pulse duration. For estimation, we also assumed $d\sigma/d\Omega\!\approx\!50\,\text{mb/sr}$ cross section for the elastic \hHe scattering at $\Ebeam\!=\!6\,\text{MeV}$ and $\tCM\!=\!90\degree$.

For the effective 1\,cm length of the target and 5\,Torr gas pressure we will detect about 100 events per pulse (per second) in the center-of-mass scattering angle range of $69\degree\!<\!\tCM\!<\!100\degree$. Thus, for $t$~min of measurements, the statistical error in value of \h beam polarization is expected to be $(\delta P/P)_\text{stat}\!\lesssim\!1.5\%\times\left(t/\text{min}\right)^{-1/2}$. A preliminary estimate of the expected systematic error for the absolute \h beam polarization measurements (assuming detailed energy scan data are available) is  $(\sigma_P/P)_\text{syst} \le 1$\%. 

\section{Summary}

The extended EBIS upgrade for heavy ion production was completed for the Run\,2023.   The next step will be integration of the \h polarizing apparatuses in the EBIS operation. The spin-rotator and the nuclear polarimeter (at 6\,MeV beam energy) construction is in progress for completion by the end of 2023. The studies of the possible depolarization effects during polarized \h gas injection and multi-step ionization process in the EBIS, the optimization of the injection valve design, and the $^3$He-cell geometry will be required to determine the maximum attainable polarization. The first EBIS operation for the polarized \hpp\ beam is planned in the Run\,2024. The expected \hpp\ ion beam intensity is about $2\times10^{11}$ ions/pulse and polarization $\ge70\%$.

\section*{Acknowledgments}

This research is funded by the DOE Research and Development Funds for the Next Generation Nuclear Physics Accelerators Facilities.


\end{document}